\documentclass[aps,prb,twocolumn,a4paper,groupedaddress,intlimits,floats,floatfix,showpacs]{revtex4-1}

\usepackage{amsmath,amssymb}
\usepackage{bm}
\usepackage[final]{graphicx}

\usepackage{epsfig}
\usepackage{color}
\usepackage{tabularx}
\usepackage{array}
\usepackage[unicode]{hyperref} 
\usepackage{multirow}

\def\nbOne{{\mathchoice {\rm 1\mskip-4mu l} {\rm 1\mskip-4mu l}
{\rm 1\mskip-4.5mu l} {\rm 1\mskip-5mu l}}}

\begin{document}
\title{Systematic computation of crystal field multiplets for X-ray core spectroscopies}
\date{\today}
\author{A. Uldry}
\author{F. Vernay}
\altaffiliation{Present address: Laboratoire PROMES (UPR-8521) \& UPVD, Perpignan, F-66860 Perpignan.} 
\author{B. Delley}
\affiliation{Condensed Matter Theory Group, Paul Scherrer Institut, 
CH-5232 Villigen PSI, Switzerland
}
\begin{abstract}
We present an approach to computing multiplets for core spectroscopies, whereby the crystal field is constructed explicitly from the positions and charges of surrounding atoms. The simplicity of the input allows the consideration of crystal fields of any symmetry, and in particular facilitates the study of spectroscopic effects arising from low symmetry environments. The interplay between polarization directions and crystal field can also be conveniently investigated. The determination of the multiplets proceeds from a Dirac density functional atomic calculation, followed by the exact diagonalization of the Coulomb, spin-orbit and crystal field interactions for the electrons in the open shells. The eigenstates are then used to simulate X-ray Absorption Spectroscopy and Resonant Inelastic X-ray Scattering spectra. In examples ranging from high symmetry down to low symmetry environment, comparisons with experiments are done with
unadjusted model parameters as well as with semi-empirically optimized ones. Furthermore, predictions for the RIXS of low-temperature MnO and for Dy in a molecular complex are proposed.
\end{abstract}

\maketitle

\section{Introduction}
X-ray Absorption Spectroscopy (XAS) and X-ray Resonant Inelastic X-ray Scattering (RIXS) measurements act as a local probe for various excitations in materials. In particular, the spectra exhibit prominently the features associated to atomic multiplets generated by open $d$ and $f$ shells. Brought by the development of synchrotron light sources, a lot of progress has been made in the last few years in RIXS, a review of which is found in Ref. \onlinecite{ament_reviewrixs}. It is now established that XAS and RIXS, especially in the soft X-ray regime\cite{ghiri_saxs}, have become the tools of choice to investigate transition metal compounds. Typical examples are the titanates\cite{abbate_91,higuchi2,higuchi,schlappa,wadati,berner_10}, the vanadates\cite{pen_YCVO,schmitt_V6O13,braico_VO2} and the cuprates in the hard\cite{hill} and soft X-ray regimes\cite{braico,schlappa_ladder}. Lately the increased resolving power of the instruments has allowed the observation of multiparticle excitations\cite{vernay_cuprates,donkov,nagao,forte_LCO,forte} arising in inelastic X-ray experiments.

 As the complexity of experiments increases and the range of materials investigated expands, it has been felt that there is the need for an easy tool relating the energy and polarization dependence of the observed spectra to the properties of crystal field atomic multiplets.

It has been known early on that the energy levels of the electrons in an open shell form multiplets under the effect of the electron-electron interaction, the spin-orbit coupling, and the crystal field\cite{condon}. While many aspects of the splitting can be discussed from a central field model and symmetry considerations alone, a quantitative approach from first principles requires the use of computers. A review of the pioneering developments in computation is given in Cowan's book \cite{cowan_book}. The early application of Cowan's approach concerned the spectra of atoms in the ultra-violet and visible range. An important further step was the inclusion of relativistic corrections to the
atomic radial functions \cite{cowan_griffin}, which was used together
with a second variation treatment of intermediate coupling
involving spin-orbit splitting and electron-electron interaction.

It was realized in the early eighties that atomic core spectroscopies
such as X-ray photoemission spectroscopy, XAS, RIXS 
and similar spectra 
can show multiplet structure signatures when open d- or f- shells 
are involved\cite{moser,thole_cowan,thole2}. An early theoretical study of atomic multiplet
splitting in a cubic crystal field has been published by van der Laan {\it et al}\cite{laan_xtal}, followed up by
de Groot {\it et al}\cite{degroot_xtal} for $3d$ transition-metal compounds.

Spectroscopic studies are typically applied to solid state compounds and therefore should in principle take into account hybridization effects. Parameterized model Hamiltonians are used to couple levels with a conduction band\cite{gunnarsson} and to include charge transfer from the selected configurations of the ligands\cite{tanaka_jo}. The inclusion of charge transfer effects in multiplet calculations is required in order to reproduce with accuracy the spectra of compounds with high ligand hole character\cite{potze}. It has however been pointed out that the simple crystal field model explains a large range of experiments\cite{degroot}. The present study will concentrate solely on the effects of the crystal field of any symmetry.

Atomic multiplet codes based on Cowan's pioneering work 
\cite{cowan,thole,missing} make intense use of symmetry considerations. Starting from the LS atomic symmetry standpoint,  the intermediate coupling arising from spin-orbit coupling is included, followed by the point group symmetries for the crystal field. The crystal field is entered in parameterized form, which depends on the point group in consideration. A somewhat simpler and more general method, where the crystal field is specified via parameters in a matrix, has also been reported\cite{mirone}. Recently, Zhang {\it et al}\cite{zhang} rationalized counter-intuitive crystal field fit parameters using a point charge model with variable ion positions.

In the present paper we focus on intra atomic electron-electron interactions
of relativistic atoms {\em in a general crystal field} and its effect on XAS and RIXS spectra. The crystal environment of the atom of interest, rather than being characterized by symmetrized matrix elements, is entered explicitly via the positions and charges of the neighboring atoms. The input is therefore straightforward and requires no prior knowledge of point group symmetry. The multiplet levels are calculated by determining the electron-electron and spin-orbit interactions, as well as the effect of the crystal field point charges for the relevant core and valence shells in a determinantal basis. The spin-orbit matrix elements and the radial wave functions are obtained from the solution of the atomic central field Dirac equation in the density functional theory formalism. The resulting eigenvalue problem is solved using standard diagonalization methods.

The paper is organized as follows. The approach to calculating multiplets 
with the point charge model defining the crystal field is presented in Sec. \ref{multiplet}. The XAS and RIXS spectra most commonly studied today are dominated by dipolar transitions caused by the interaction of polarized light with the relevant core and valence electrons of the atom. A brief reminder of the standard theoretical treatment for XAS and RIXS and the selection rules are given is Sec. \ref{spectroscopy}. The application of the present approach to both well known high symmetry and less familiar low symmetry cases will be demonstrated is Sec. \ref{xas} for XAS and in Sec. \ref{rixs} for RIXS. Summary, conclusions and outlook follow in Sec. \ref{conclusions}.

\section{Atomic multiplets for general crystal field\label{multiplet}}
The general theory of atomic multiplet is described extensively in 
Refs. \onlinecite{condon,cowan_book,degroot,balhausen}. Within the point charges model, the relevant interactions for the electrons in the open shells are the intra-atomic electrostatic interactions, the spin-orbit coupling and the effect of the crystal field. Considering $n$ electrons in the open shells of an atom and $N_{\text{ions}}$ point charges $Q_m$ at position $\mathbf{R}_m$ from the atom, the multiplet Hamiltonian 
is given by 
\begin{equation}\label{mult_hamil}
\mathcal{H_{\text{mult}}}=
\sum_{i,j}^{n} \frac{e^2}{| {\bf r}_i-{\bf r}_j |} +
\sum_i^{n} \varepsilon_i
+\mathcal{V}_{\text{xtal}}
\end{equation}
The third term $\mathcal{V}_{\text{xtal}}$ is the crystal field interaction in the crystal field potential $V_{\text{xtal}}$
\begin{equation}\label{Vxtal}
V_{\text{xtal}}({\bf r}) =\sum_{m=1}^{N_{\text{ions}}} 
\frac{Q_m}{| {\bf r}-{\bf R}_m |}
\end{equation}
The first term in Eq. \ref{mult_hamil} is the electron-electron Coulomb interaction. The second is a single electron operator accounting for the kinetic energy and spin-orbit interaction; this term will be drawn from the solution of the atomic, central field Dirac equation, a point that will be discussed further below in this section. A typical case involving a transition metal will require considering the multiplets arising from opening a hole in the $2p$ shell and promoting an electron in an incomplete $3d$ shell, so that all the electrons in $2p$ and in the $3d$ shells will be included in Eq. \ref{mult_hamil}.

The bulk of this part of the paper addresses some aspects of the implementation of the three terms of Eq. \ref{mult_hamil}. The Hilbert space is presented first. Then, starting with the central field atomic equation, we discuss the spin-orbit coupling, the crystal field, and the effect of ionization. A comparison with other existing approaches is conducted at the end the section.

\subsection{The Hilbert space}
The focus of this paper lies in calculating multiplets involved in soft X-ray absorption and emission processes. The wave functions are antisymmetrized products of the spin-orbitals of the electrons in the atomic open shells, so the dimension of the Hilbert space is, in the present model,
\begin{equation}  
{\mathcal N}(l,k) = \frac{[2(2l+1)]!}{[2(2l+1)-k]!\ k!}
\end{equation}
for a shell labeled by the quantum 
numbers ($n,\ l$) and containing $k$ electrons.
For instance 5 electrons in a $p$-shell gives ${\mathcal N}(1,5)$=6 determinants, 
 2 electrons in a $d$-shell gives ${\mathcal N}(2,2)=45$ determinants.  
If more than one shell is opened, each shell being independent, the overall 
size is given by the product ${\mathcal N}=\prod_{i}{\mathcal N}_i$. The case of transition metals can be handled easily on today's standard desktop computers. The largest Hilbert spaces in present day soft X-ray applications are found in the case of lanthanides, for which $d\to f$ transitions with an open $f$ shell occur. An example could be Gd$^{3+}$ in the ground state configuration $3d^9 4f^7$, 
with a final state matrix dimension of 34320. The generation and diagonalization of this size of matrices remain however still tractable on contemporary architecture with sufficient memory.

\subsection{Relativistic electronic orbitals and energies for a single atom}
Computing atomic multiplet levels requires central field model orbitals and energies. Previous atomic multiplet calculations, including  the more recent approaches\cite{mirone}, are usually based on the non-relativistic Schr\"odinger equation. In this work the Dirac equation has been preferred,  since it naturally includes spin-orbit coupling effects. 

In the absence of crystal field, the Dirac equation is written in a formal way as follows:
\begin{equation}
{\mathcal H_D}\Psi_i=\varepsilon_i\Psi_i\ \ ;\ \
{\mathcal H_D}= c\bm \alpha\cdot \bm p+\bm \beta mc^2+V(r)
\end{equation}
\begin{equation}
{\bm\alpha}=
\begin{pmatrix}
0               &      {\bm\sigma}\\
{\bm\sigma}     &          0      \\
\end{pmatrix}\ ;\ 
{\bm\beta}=
\begin{pmatrix}
\nbOne_{2}   &   0\\
0             &-\nbOne_{2} \\
\end{pmatrix}
\end{equation} 
where $\bm\sigma$ are the Pauli matrices and $V(r)$ is a spherically 
symmetric field which can be decomposed as follows:
\begin{equation}
V(r)=-\frac{{
Z}}{r} +V_s(r)+V_{xc}(r)\ ;
\end{equation}
The first term is the electrostatic attraction by the nuclear charge $Z$ and
$V_S$ is the static potential: 
\begin{equation}\label{Vs}
V_s(r)=\frac{1}{r}\int_0^r\rho(x)4\pi x^2 dx 
+ \int_r^\infty\frac{\rho(x)}{x}4\pi x^2 dx.
\end{equation}

In the multiplet code developed for the present work the Dirac equation is solved for the neutral atom (see Sec. \ref{ioniz}), within the density functional theory\cite{kohn} and using a local density functional approximation $V_{xc}(r)$. The radial wavefunctions and eigenvalues $\varepsilon_i$ will be used to compute the matrix elements of Eq. \ref{mult_hamil}. For light elements, the non-relativistic limit of the radial wave functions can be retrieved by performing a weighted average of the radial term over the shell. For the examples shown in this work the crystal field matrix elements have also been calculated in this limit.

For the application to XAS and RIXS the Dirac equation has to be solved twice: once in the ground state and once with an electron moved from a core shell to a valence shell.

\subsection{Spin-orbit coupling and Coulomb term \label{ioniz}}
As mentioned above, the eigenvalues $\varepsilon_i$ of the Dirac equation are used to build a diagonal operator that, by default, accounts for the effect of spin-orbit coupling (second term in Eq. \ref{mult_hamil}). All the electrons of the participating core and valence shells contribute to the spin-orbit operator.

Such an operator includes the kinetic energy, but also contains the contribution of the central symmetric part of the electron-electron interaction, which in principle is already addressed by the first term (Coulomb term) in Eq. \ref{mult_hamil}. 

Double counting can be partially prevented by considering the following standard expansion that applies to the Coulomb term in Eq. \ref{mult_hamil},
\begin{equation}\label{rexpand}
\frac{1}{|\bm{r_1 - r_2}|} =\sum_{k=0}^{\infty}\frac{4\pi}{2k+1}
\sum_{m=-k}^{k}
\frac{r_1^k}{r_2^{k+1}}
Y_k^m(\theta_1,\phi_1)^\ast Y_k^m(\theta_2,\phi_2)
\end{equation}
with $r_1<r_2$.
Replacing the sum 
$\sum_{k=0}^\infty$ with $\sum_{k=1}^\infty$ in the expansion of Eq. \ref{rexpand} effectively removes a spherically symmetric term corresponding to $V_s$ in Eq. \ref{Vs}, and part of the exchange interaction $V_{xc}$.

Alternatively, and this is the approach used in all the examples presented in this paper, the radial term in the matrix element of the Coulomb interaction is averaged over the shell, thereby effectively removing unwanted contributions.

It will be seen from comparisons with X-ray spectroscopic experiments in Sec. \ref{xas} and \ref{rixs} that this straightforward approach to spin-orbit coupling delivers a good first-principle prediction of the 
L$_3$-L$_2$ edge splitting.

\subsection{Crystal Field as point charge ligands \label{xtal_intro}}
Unlike group-theoretical approaches to calculating multiplet levels, the present work accounts for crystal field splitting by explicitly including the electrostatic interactions of neighboring point charges with the open shells of the atom. The crystal field is then computed together with the direct and exchange Coulomb interactions, and the spin-orbit coupling (Eq. \ref{mult_hamil}). 

The advantage of treating the crystal field as point charge ligands is two-fold: {\it i)} the position of each ion is given as input in the program without having to resort to symmetry and 
{\it ii)} the crystal field arising from charges at experimental 
atomic positions leads to a reasonable prediction with often sufficient 
accuracy for interpretation without using a fitting parameter. \\
The purpose of the charges is to recreate the proper symmetry of the crystal. It is natural to take as a starting point the oxidation state, or the formal charges, the latter reflecting the level of band filling in insulators. Usually however, either fail to reproduce correctly the actual charge distribution in the solid, and it would be legitimate to treat the charges assigned to equivalent atoms as fitting parameters. This exercise has not been carried through in the examples discussed in this work.

\begin{figure}[]
\includegraphics*[width=8cm,angle=0]{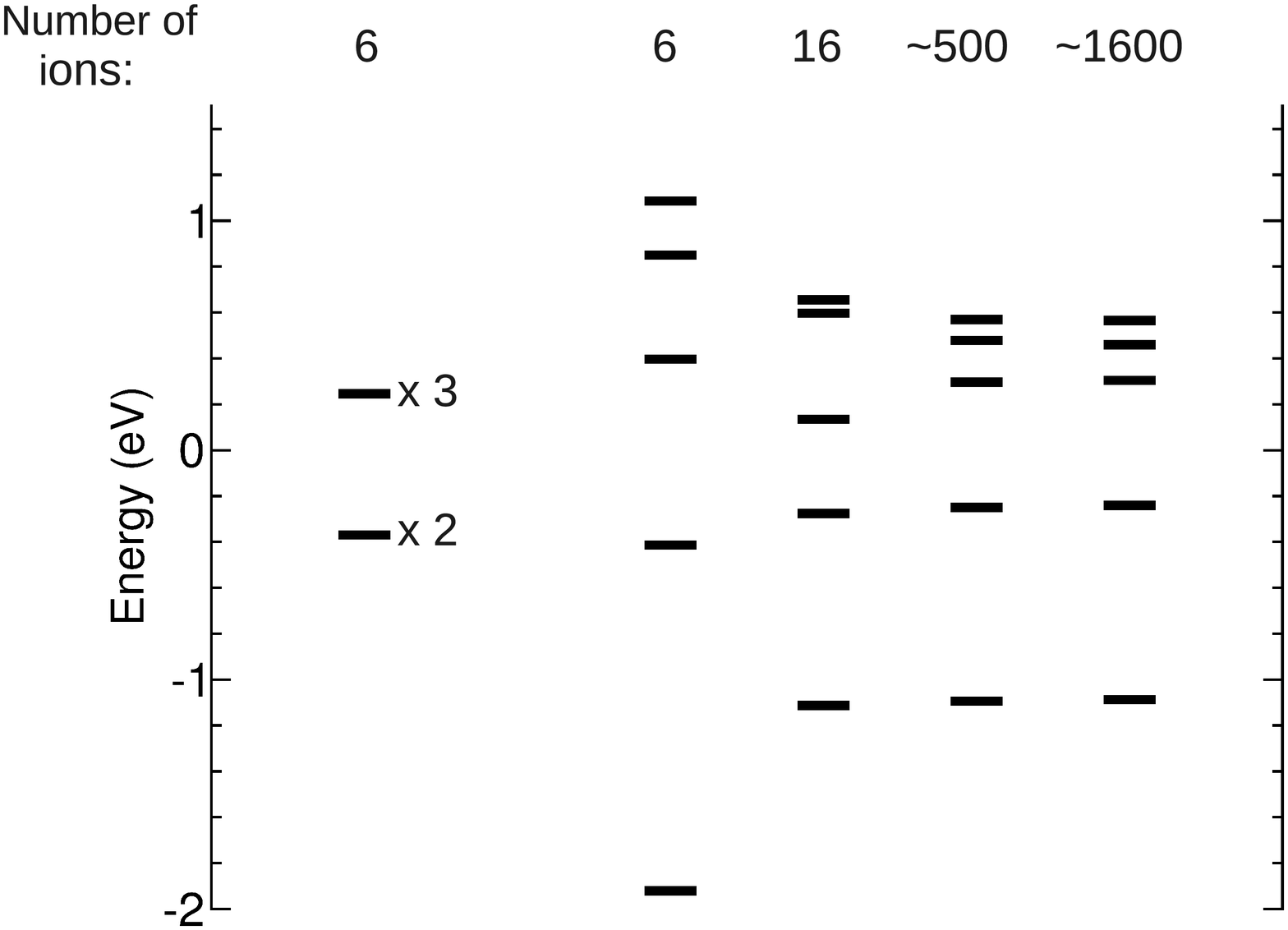}
\caption{Multiplet levels of Cu$^{2+}$ in CuO calculated with the present code in the NR limit. Column 1: cubic octahedral environment (6 O atoms). Columns 2-5: monoclinically distorted environment built up of the 6, 16, $\approx$ 500 and $\approx$ 1600 closest ions, respectively.\label{CuOsplit}}
\end{figure}

An example of input is given in Table \ref{inputlig} for the monoclinic tenorite (CuO), where the Cu has 6 nearest neighbors forming a distorted octahedral environment. 
\begin{table}
\caption{\label{inputlig} Positions and charges of the crystal field for Cu$^{2+}$ in monoclinic tenorite CuO. The oxygen coordinates x,y,z and distance r to Cu are given in \AA, the charges in units of e. For comparison, the cubic polymorph of CuO has 6 oxygen atoms at a distance of r=2.12 \AA.}
\begin{tabular}{c|c|c|c|c}
   x  & y & z & r & charge \\
  \hline
  \hline
 1.3304	& 0.5617 & -1.2461 & 1.910 & -2.0 \\
0.9001	& 1.1298 & 1.2836 & 1.930 &-2.0\\
-1.0143	& -1.1517 & -1.2461 & 1.980 & -2.0\\
-1.4446	& -0.5836 & 1.2836 & 2.020 & -2.0\\
0.9001	& -2.2970 & 1.2836 & 2.780 & -2.0\\
-1.0143	& 2.2751 & -1.2461 & 2.790 & -2.0
\end{tabular}
\end{table}
The positions of the neighbor O atoms are taken from the Inorganic Crystal Structures Database (ICSD) entry number 69758. A formal charge of -2e was allocated to each O atom.
The ground state configuration of Cu$^{2+}$ is $2p^63d^9$. This information, together with the elements of Table \ref{inputlig}, are the only input required in the code. In that particular case, and in the non-relativistic (NR) limit, the splitting of levels originates solely from the crystal field. The output of the code is represented graphically in Fig. \ref{CuOsplit}. 
The first column on the left shows the levels for Cu$^{2+}$ in octahedral symmetry. This structure exists and is a polymorph of CuO in a cubic phase (ICSD 61323), where the six O atoms are forming an octahedron at a distance of 2.12 \AA $\,$ to the Cu. If the crystal field of Table \ref{inputlig} is considered instead (column 2), the energy levels originally ascribed to the $t_{2g}$ (3-fold degenerate) and $e_g$ (2-fold degenerate) orbitals in the cubic case split into three, respectively two further levels.\\
The convergence with increasing numbers of ions as ligands is illustrated by the columns 2-5 in Fig. \ref{CuOsplit}, with respectively 6, 16, $\approx$ 500 and $\approx$ 1600 closest ions considered. While the explicit symmetry breaking induced by the crystal field is already accounted for by the 6 closest oxygen atoms, the addition of the second shell of ions (16 closest ions) is necessary to reach a better qualitative agreement with the limit. In the particular example of monoclinic CuO, the convergence is achieved for most practical purposes with about 500 ions.

\subsection{Semi-empirical parameters}\label{ionization}
The assumption behind the crystal field model is that the atom under consideration is surrounded by point charges and is itself in an particular ionized state. The ionization has a major impact on the radial part of the wave functions of an atom. Moreover, the radial wave functions will be affected by the bonding environment of the atom. The latter consideration is the origin of the introduction of semi-empirical parameters. As the radial functions enter in both the calculation of the crystal field and the Coulomb term, the sensitivity of these terms to the formal ionization state is addressed below.

\begin{figure}
\includegraphics*[width=8cm,angle=0]{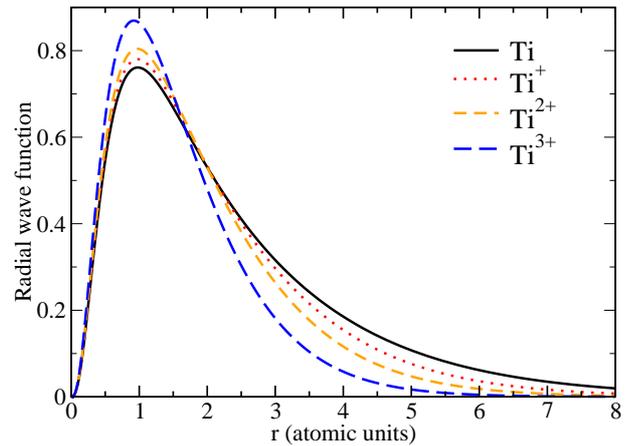}
\caption{(Color online) 
Radial part of the $d$-wave function for Ti as 
a function of $r$ in atomic units: the tail of the wave function shrinks 
with increasing ionization.}\label{radial}
\end{figure}

An atom give rises to different radial wave functions according to the degree of ionicity. The reduced screening felt by the remaining electrons of a positively charged atom leads to the contraction of the radial wave functions around the nucleus. As illustrated in Fig. \ref{radial} for the case of Ti, the expectation value $\langle r \rangle$ in the Ti$^{3+}$ ion varies by about $30\%$ compared to the neutral titanium ion.
An estimation of the effect of the contraction of the radial function can be given as follows. For an open $d$ shell in a non-relativistic limit, the only contribution to level splitting comes from the anisotropic densities with a purely $k$ = 4 multipolar character. Eq. \ref{rexpand} implies that the crystal field splitting scales with the power of the anisotropic density. As the radial wave function becomes sharper with increasing ionization, the crystal field splitting for a shell of angular momentum $l$ scales as $\langle r^{+2l} \rangle $. On the other hand, the electron-electron interaction scales like $\langle r^{-1} \rangle$. As a consequence, if the wave function does actually extend further than anticipated, the crystal field strength would be under-evaluated, whereas the Coulomb interaction would be over-evaluated. \\

A comparison with experimental spectra indicates that a better correspondence is achieved if the radial functions of the {\it neutral} atom with integral ionic charge number is used, rather than those of the positively charged ion. Moreover, self-consistent band-structure calculations suggest that the charge density around atoms in a solid is close to that of neutral atoms. All the examples presented in this paper have therefore been obtained with radial functions calculated from the neutral atom.

Further comparisons with experiments show, unsurprisingly, that an even better match can be made if the matrix elements of the Coulomb interaction, the spin-orbit coupling and the crystal field are allowed to vary. A scaling of these elements can be seen as a way to adjust the radial functions to the hybridization environment encountered in the solid. As will be seen in examples below, it is usually necessary to scale the crystal field up, and the coulomb interaction down. From the discussion in the paragraph above, this suggests that the neutral atom radial wave functions still fall off too rapidly.\\

An empirical correction to the ionic model is therefore done in our code {\it via} the parameters S$_\text{coul}$, S$_\text{xtal}$ and S$_\text{soc}$, which scale the respective terms in the Hamiltonian (Eq. \ref{mult_hamil}). The crystal field scaling may also incorporate a global adjustment to the chosen charges. \\

We will call ``raw'' the multiplet results or spectra calculated using formal charges, respectively oxidation states, and crystallographic neighbor positions, all scaling parameters being set to unity. \\

The model of a single ion in an ionic environment ignores charge transfer from ligands, band effects and collective excitations. It has nonetheless been successfully applied to the calculation of multiplet-dependent spectral features\cite{ghiri}. Ghiringhelli {\it et al} compared the crystal field model to the single impurity Anderson model for a prototype transition metal oxide, MnO, in a cubic environment. They found that both models agree well in that part of the spectra determined by local effects. The multiplet approaches\cite{laan_xtal,degroot_xtal} based on Cowan's work\cite{cowan_book} use Hartree-Fock parameters (which include relativistic corrections). These values are usually scaled down by 70 to 80\% to compensate for the neglect of configuration interactions in the Hartree-Fock theory. The effective crystal field parameters are adjusted to experiments. The present approach makes use of the DFT-LDA results of the relativistic Dirac equation; in principle it delivers the multiplet structure without the need of parameterization; if fitting to experiments is required, the scaling parameters and possibly the charges can be adjusted.

\subsection{Results and comparison\label{mult_results}}
In the presence of a cubic crystal field and in the non-relativistic limit, the energy levels of one electron in a $d$ atomic orbital split into a four-fold degenerate $e_g$ and a six-fold degenerate $t_{2g}$ orbital (including spins). In this case, the cubic crystal field splitting is well defined as the energy difference between the $e_g$  and the $t_{2g}$ levels. This value is also often named $10D_q$ in the literature, and is taken as an effective parameter fitted to experiments\cite{degroot_xtal}. In Ref. \onlinecite{degroot_xtal}, the authors estimated that for transition metals, the crystal field splitting would typically vary between 0 and 2.5 eV. They point out that the splitting found by examining optical spectroscopy experiments correspond to the final state splitting, which can in principle differ from that of the ground state initial value. 

As the present work deals with low as well as high symmetry, a characteristic crystal field splitting (CXS) is defined, which expresses the strength of the crystal field splitting. For cubic symmetry the CXS is equivalent to the cubic crystal field splitting. In the present approach, the CXS can be easily determined for a cubic field by evaluating the central atom in a $d^1$ or $d^9$ configuration in the non-relativistic limit. 
In the example of Cu in cubic geometry, as can be seen in Fig. \ref{CuOsplit}, the unscaled, raw result of the calculation gives CXS=0.6 eV. In the monoclinic structure the strength of the crystal field can be taken as the energy difference between the barycenter of the two lowest levels and that of the 3 highest levels, which gives 1.9 eV. Alternatively, one can take the CXS as the difference between the highest of the two lowest and the average of three highest levels, which amounts to 0.7 eV (for the $\approx$ 500 ions case).  

A well studied example is Ti$^{3+}$ in LaTiO$_3$, in its $2p^63d^1$ ground state configuration. Once thought of as a perfect cubic perovskite\cite{kestigian}, LaTiO$_3$ has been characterized a number of times; a later powder diffraction study\cite{cwik} detected small orthorhombic distortions to the octahedral environment. The multiplet results of our code for LaTiO$_3$ in the idealized cubic structure\cite{kestigian} (ICSD 28908) and with orthorhombic distortions\cite{cwik} (ICSD 98414) is presented in Table \ref{LaTiO3mult}. 
\begin{table}
\caption{\label{LaTiO3mult}Multiplets of the ground state $2p^63d^1$ of LaTiO$_3$.  Eigenvalues are given in eV for unscaled (raw) contributions to the Hamiltonian (Eq. \ref{mult_hamil}), and with scaling determined from fitting to XAS data\cite{haverkort}, as explained in the text. The degeneracy deg (including spin) is indicated for the cubic\cite{kestigian} and orthorhombic\cite{cwik} ligands field.}
\begin{tabular}{c|c c|c c c}
   & \multicolumn{2}{c|}{Cubic} & \multicolumn{3}{c}{Orthorhombic}\\	
   & deg & raw & deg & raw & scaled\\
  \hline
\multirow{5}{*}{\begin{tabular}{c} 
Crystal field \\ (NR limit) \end{tabular}} & 
\multirow{5}{*}{\begin{tabular}{c} 6 \\ 4 \end{tabular}} &
\multirow{5}{*}{\begin{tabular}{c} -1.083 \\  1.624 \end{tabular}}		        &	
                                                                2	&	-1.200	&	-1.500	\\
                               &		&	&	2	&	-0.849	&	-1.062	\\
	                       &		&     	&	2	&	-0.735	&	-0.919	\\
	                       &		&      	&	2	&	1.327	&	1.659	\\
	                       &		&	&	2	&	1.458	&	1.822	\\
\hline
\multirow{2}{*}{Spin-orbit}	&	4	&	-0.027	&	4	&	-0.027	&	-0.025	\\
	                        &	6	&	0.018	&	6	&	0.018	&	0.017	\\
\hline
\multirow{5}{*}{\begin{tabular}{c} Crystal field \\ + Spin-orbit \end{tabular}    }	&		&		&
                                                                   	2	&	-1.200	&	-1.500	\\
	                        &	4	&	-1.092	&	2	&	-0.850	&	-1.062	\\
	                        &	2	&	-1.065	&	2	&	-0.735	&	-0.919	\\
	                        &	4	&	1.624	&	2	&	1.327	&	1.659	\\
	                        &		&		&	2	&	1.458	&	1.822	\\
 \hline
\end{tabular}
\end{table}
The formal charges La$^{3+}$, Ti$^{3+}$ and O$^{2-}$ were allocated to the ions.
The raw, ``first-principles'' eigenvalues are obtained with, as sole input, the specification of Ti as the atom under consideration, its ground state configuration $2p^63d^1$ and the crystal field as described above. For both structures all charges within 20 \AA $\,$ of the Ti atom were included. The first row is obtained by considering only the Coulomb interactions and the crystal field in the NR limit. As expected, the cubic crystal field causes a $t_{2g}$-$e_g$ split, giving CXS=2.7 eV. The effect of the spin-orbit coupling, without the crystal field (second row in Table \ref{LaTiO3mult}), is to split levels by 45 meV. The combination of cubic crystal field and spin-orbit coupling has the effect of splitting the $t_{2g}$ level by 27 meV (third row in the Table). With the orthorhombic field (column labeled Orthorhombic raw) all the levels are already split by the crystal field in the NR case. The inclusion of spin-orbit interactions (third row in the figure) has only a minute effect on the levels. Taking CXS as the energy difference between the highest negative value and the average of the positive value, we find CXS=2.1 eV in the orthorhombic case. \\

 The scaled results for LaTiO$_3$ are shown in Table \ref{LaTiO3mult} along the original, raw values for the orthorhombic geometry. The scaling parameters were obtained by fitting to XAS experiments\cite{haverkort} (see results in Sec. \ref{xas}), which gave S$_\text{coul}$=0.7, S$_\text{xtal}$=1.25 and S$_\text{soc}$=0.915. The corrected CXS now has a value of 2.7 eV (2.1 eV unscaled). In the scaled case the splitting of the $t_{2g}$ level is of about 0.14 and 0.4 eV. These values agree well with the measurements and LDA-based, Wannier function projection calculations of Ref. \onlinecite{haverkort}.

\section{XAS and RIXS theory \label{spectroscopy}}
XAS and RIXS involve processes whereby an X-ray photon is absorbed by a core electron, which is promoted into a valence shell. The resulting core-hole and valence configuration, in general, gives rise to a multiplet structure. In the present work the ground state and core-hole state are determined using the multiplet Hamiltonian (Eq. \ref{mult_hamil}) as sketched in Sec. \ref{multiplet}.

XAS measures the decay products of the core-hole\cite{ament_reviewrixs}. In RIXS measurements the system returns to the ground state by emitting a photon. The standard theoretical formulation for the calculation of the signal intensity of XAS and RIXS is stated below, as well as the dipolar approximation used in both cases. Lastly, a brief reminder of the polarization effects and selection rules is given.
\subsection{XAS intensity formula}
The XAS absorption spectra is simulated the usual way\cite{degroot} with the Fermi Golden rule
\begin{equation}\label{abs1}
I(\omega)\propto\sum_i|\langle\psi_i|\hat{{\mathcal O}}|\psi_0\rangle|^2
\delta(\hbar\omega+E_0-E_i)
\end{equation}
$\hbar\omega$ is the energy of the absorbed photon.
$|\psi_0\rangle$ refers to the ground-state and $ E_0$ to its eigenvalue, which is often degenerate.
$|\psi_i\rangle$ and $E_i$ are respectively all the possible eigenvectors and energies of the excited, core-hole state. The transition operator $\hat{\mathcal O}$ is treated in the dipolar approximation, briefly exposed in Sec. \ref{subsec:dipolar}, and depends on the polarization of the incoming light $\epsilon_{in}$ (Sec. \ref{subsec:polariz}).

Intrinsic effects as well as the finite experimental resolution leads inevitably to the broadening of the sharp delta peak of Eq. \ref{abs1}. 
The spectral broadening here is approximated by a Lorentzian. Eq. \ref{abs1} becomes:
\begin{equation}
I(\omega)\propto\sum_i|\langle\psi_i|\hat{{\mathcal O}}|\psi_0\rangle|^2
\frac{\Gamma_i/\pi}{(\hbar\omega+E_0-E_i)^2+\Gamma_i^2}
\end{equation}  
An intrinsic contribution to the Lorentzian broadening comes from the finite lifetime of the core-hole state. A typical $L_{2,3}$ core-hole linewidth for transition metals varies between 0.2 and 0.4 eV\cite{pease}. The different level of hybridizations of the states involved may however be reflected in a variation of the vibrational and dispersional broadening of the peaks.\cite{degroot_broad}.

If $\Gamma_i \equiv \Gamma$ is taken as a constant, the relative intensities of the simulated peaks may differ from that observed in experiments. A Gaussian experimental broadening can be applied as well if a better fit to experiment is sought; this has not been done in the examples shown in this work.

The multiplet levels of the excited core-hole state are given with respect to the energy difference between the total energy of the single neutral atom in the ground state and in the excited state, which in the present case is determined within the LDA-DFT approximation. The absolute energy position of the multiplets is not determined with sufficient accuracy and, compared to experiments, tends to shift the peaks to a lower energy by an amount of the order of the percent. It is usual to align the calculated spectrum to the measured one, and this has been systematically done in this work for the XAS spectra.

\subsection{RIXS intensity formula}
RIXS processes involve the creation and subsequent radiative deexcitation of a core hole. With the incoming photon energy $\hbar\omega_{in}$ close to an absorption edge, the RIXS intensity is approximated, using the Kramers-Heisenberg formula\cite{kramers_heisenberg},  by the resonant term
\begin{eqnarray}\label{Eq:KH}
I(\Omega,\omega_{in})=
 \displaystyle\sum_f \left|
\displaystyle\sum_i \frac{\langle \psi_f|{\mathcal O'}^\dagger|\psi_i\rangle
\langle \psi_i|{\mathcal O}|\psi_0\rangle}
{E_i-E_0-\hbar\omega_{in}-i\Gamma_i}\right|^2 \nonumber \\
\times \delta\left(E_f-E_0-\hbar\Omega\right) 
\end{eqnarray}
Details on the second order perturbation theory leading to Eq. \ref{Eq:KH} can be found in the review Ref. \onlinecite{ament_reviewrixs}. $\hbar\Omega$ is the energy loss, i.e. the energy difference between the incoming and outgoing photon.
During the first stage of the process a core-electron is promoted from the ground-state $|\psi_0\rangle$ of energy $ E_0$ into an excited, core-hole intermediate state $|\psi_i\rangle$ of energy $E_i$. The probability of 
this transition depends on the polarization of the incoming light and is determined by the dipolar operator 
$\mathcal O$. The deexcitation from the intermediate to a final state $|\psi_f\rangle$ of energy $E_f$ occurs via the emission of a photon of energy  $\hbar\omega_f$ and depends on the polarization of the outgoing photon, as given by the operator $\mathcal O'$ (see Sec. \ref{subsec:polariz}). The core-hole lifetime $\Gamma_i$, as above in the XAS case, is often taken as a constant. The $\delta$ function in Eq. \ref{Eq:KH} is also broadened by the finite lifetime of the final state $|\psi_f\rangle$ and experimental resolution. A Lorentzian function is usually used {\it in lieu} of the delta function, as is the case for the present work.

The experimental RIXS spectra also contain broad fluorescence lines
which mainly come from the emission from intermediate states with 
finite lifetimes which are not taken into account in the multiplet 
calculation. Non-local contributions are also ignored in this study for the present.

\subsection{Dipolar Approximation \label{subsec:dipolar}}
The incoming photon is defined by its polarization ${\bm \epsilon}$ and its 
wavevector ${\bf k}$. After expanding the vector potential in plane waves, and in the limit ${\bf k}\cdot{\bf r}\ll 1$,  the matrix elements in Eq. \ref{abs1} and \ref{Eq:KH} can be obtained as
\begin{equation}
\langle\psi_i|{\bf\epsilon}\cdot{\bf p}e^{i{\bf k}\cdot{\bf r}}|\psi_0\rangle
\sim
\langle\psi_i|{\bf\epsilon}\cdot{\bf p}
\left(1+i {\bf k}\cdot{\bf r} - ...\right)|\psi_0\rangle
\end{equation}
The dipole approximation consists in keeping only the first term, such that 
the operator is given by
\begin{equation}
\hat{\mathcal O}= {\bf\epsilon}\cdot{\bf p}={\bf\epsilon}\cdot
[{\bf r},{\mathcal H}]
\propto {\bf\epsilon}\cdot{\bf r}
\end{equation}
Using an expansion on the $Y_1^m$, we finally obtain
\begin{equation}\label{dipolop}
\hat{\mathcal O}\propto r\left(\epsilon_1 Y_1^1+\epsilon_0 Y_1^0
+\epsilon_{-1} Y_1^{-1}\right)
\end{equation}
where the coefficients $\epsilon_i$ represent the projection of the 
polarization vectors on the $Y_1^m$ basis.  

At the dipolar approximation level the selection rules 
for the transition between states are the following:
\begin{equation}
\Delta l=\pm 1\ ;\ \Delta s=0\ ; \Delta J=0,\pm 1 ;\ ...
\end{equation}
In the present code these rules are automatically included 
through the computation of Gaunt coefficients formed by combining the spherical harmonics of Eq. \ref{dipolop} and that of the wave functions of the matrix elements in Eq. \ref{abs1} and \ref{Eq:KH}. The weights of the Gaunt coefficients are given by the polarization projections. The polarization has therefore a discriminating effect on the transitions.

\subsection{Polarization effects\label{subsec:polariz}}
 The approach presented in this work is well suited to the study of polarization effects in low symmetry. The polarization of the photon is given in the same frame of reference as the ions forming the crystal field, and therefore any possible geometries can be analyzed quantitatively.
\begin{figure}[]
\includegraphics*[width=5cm,angle=0]{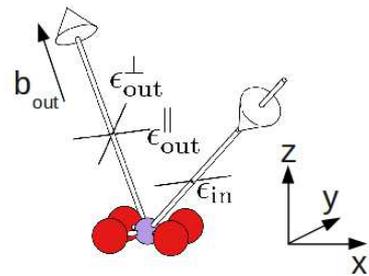}
\caption{(Color online) Polarization directions with respect to the crystal field: an incoming beam of photons of polarization $\bm \epsilon_{\text{in}}$ excites the central atom (small purple ball). The outgoing photons that are detected along a direction $\bm b_\text{out}$ can have any polarizations perpendicular to it. As an example here $\bm \epsilon^{\parallel}_{\text{out}}$ and $\bm \epsilon^{\perp}_{\text{out}}$ are also parallel, respectively perpendicular, to $\bm \epsilon_{\text{in}}$. The four ions (large red balls) making up the planar environment of the central atom are described in the same reference frame $({\bf x,y,z})$ as the polarization and beam directions.}\label{fig:polariz}
\end{figure}
An example is shown in Fig. \ref{fig:polariz} for an atom in a planar environment. The 4 ions forming the crystal field are expressed in the local $({\bf x,y,z})$ basis, and so is the polarization of the incoming and outgoing photon. In most RIXS experiments the polarization of the outgoing photons is currently not recorded, and only the detection direction $\bm b_\text{out}$ of the beam is known. The signal is simulated by the incoherent superposition of intensities contributed by the polarizations orthogonal to the detection direction. A similar superposition is done in the case of unpolarized photons, or experiments done on powder samples.

As pointed out above, the weights of the Gaunt coefficients are determined by the polarization. In the most general case different polarizations will suppress or enhance other transitions within the dipole approximation. Examples of the effect of polarization directions will be given in the Sec. \ref{xas} and \ref{rixs}.

\section{XAS spectra\label{xas}}
In this section several XAS spectra simulated with the present code are compared with experiments. First the familiar high symmetry case of the SrTiO$_3$ is reproduced. Then the spectrum of LaTiO$_3$ with cubic and with orthorhombic distortions is presented. Finally, the monoclinic epidote is discussed. \\

\subsection{Cubic SrTiO$_3$}
\begin{figure}
\includegraphics*[width=8cm,angle=0]{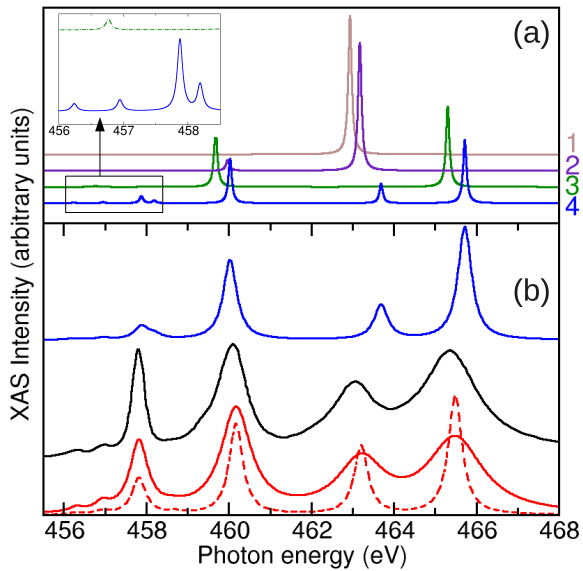}
\caption{(Color online) 
XAS spectra for Ti$^{4+}$ in SrTiO$_3$ in the $2p^6 3d^0$ ground state configuration. (a) Raw results of the multiplet code. Lines from top to bottom: Coulomb interaction in the NR limit (1, brown), then adding crystal field (2, purple), then for Coulomb interactions and spin-orbit coupling (3, green), and finally for Coulomb, spin-orbit and crystal field (4, blue). (b) Top line (blue): raw results of the multiplet code with all contributions and a constant core-hole lifetime; middle line (black): experimental result of 
J. Schlappa {\it et al.}\cite{schlappa}; lower curves (red): results with rescaling of all contributions as explained in the text. The dashed line is produced with a constant core-hole broadening, the full line with a broadening increasing with increasing energy. \label{xas_SrTiO3} }
\end{figure}

SrTiO$_3$ is an example of cubic perovskite with formal charges Ti$^{4+}$, Sr$^{2+}$ and Ti$^{2-}$. The XAS spectrum at the titanium L-edge will therefore involve the transitions $2p^63d^0$ to $2p^53d^1$. While no multiplet structures appear in the ground state, the Hilbert space for the core-hole state involves 60 determinants. The multiplet levels reflects the non-trivial interplay between the intra-atomic Coulomb interaction and spin-orbit, and the crystal field, as is illustrated in Fig. \ref{xas_SrTiO3}(a), a raw result of the code.
An artificially small core-hole width of 0.05 has been chosen in order to exhibit the positions of the peak more clearly. The cubic  crystal field is such that the 6 oxygen neighbors are at a distance of $d=1.95$ \AA. In the non-relativistic limit the pure Coulomb peak (curve 1, brown) splits in two (curve 2, purple) upon the addition of the crystal field. In the absence of crystal field, the spin-orbit coupling splits the peak in three (curve 3, green), as can be seen with the help of the zoom in the inset in the figure. The bottom line (curve 4, blue) is the result after including all contributions. Fig. \ref{xas_SrTiO3}(a) has been produced from the raw results of the code without applying any rescaling.
Fig. \ref{xas_SrTiO3}(b) compares the XAS intensity calculations with the experimental measurements (middle curve, black) taken from Schlappa {\it et al}\cite{schlappa}. This measurement is in very good agreement with that of other authors\cite{abbate_91,berner_10}.
The top curve (blue) is the raw result of the code; it is the same as the bottom curve of (a), but with a constant $\Gamma_i$ broadening of 0.2 eV, corresponding to a realistic intrinsic core-hole lifetime value. The bottom curves (red) are the result of rescaling with S$_\text{coul}$=0.85, S$_\text{xtal}$=1.3 and S$_\text{soc}$=0.93. Keeping the core-hole broadening constant at 0.2 eV produces the dashed line. In this particular case the rescaling of the spin-orbit interaction has the effect of bringing the L$_3$ and L$_2$ peaks slightly closer together. The relative heights of the peaks is sensitive to the scaling of the intra-atomic Coulomb interaction. The energy difference between the two L$_3$ peaks (and the two L$_3$ peaks) depends predominantly on the crystal field strength and therefore determines the rescaling of the crystal field. The agreement with experiment can be further improved by modeling the energy dependence of $\Gamma_i$. The inclusion of a theoretical approach for the core-hole lifetime is beyond the scope of this paper, but the broadening can be empirically matched to the experiment. The full  bottom line (red) in Fig. \ref{xas_SrTiO3}(b) was obtained by increasing $\Gamma_i$ from 0.2 eV to 0.8 eV linearly with $E_i$ between 457 and 465 eV.

Our results compare very well with simulated spectra produced by multiplet approaches where the crystal field is introduced by symmetry group parameters and strength\cite{abbate_91}.

\subsection{Cubic and orthorhombic LaTiO$_3$}
\begin{figure}
\includegraphics*[width=8cm,angle=0]{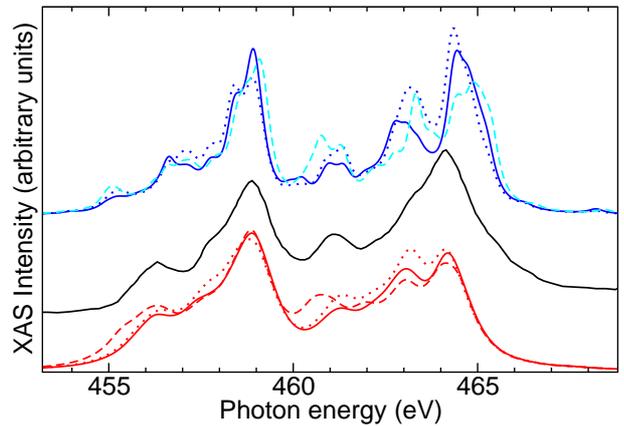}
\caption{(Color online) 
XAS spectra for Ti$^{3+}$ in LaTiO$_3$ in the $2p^6 3d^1$ ground state. Top (blue) lines: raw multiplet results for cubic crystal field (dashed), orthorhombic crystal field with polarization A (full) and B (dotted line) as described in the text. Middle (black) line: experiment as reported in Haverkort {\it et al}\cite{haverkort}. Lower (red) lines: rescaled multiplet results for the cubic crystal field (dashed), and for the orthorhombic crystal field with polarization A (full) and B (dotted line).\label{xas_LaTiO3} }
\end{figure}

The ground state multiplets for the Mott-Hubbard insulator LaTiO$_3$\cite{imada} has already been discussed in Sec. \ref{mult_results}. The resulting XAS spectra are plotted in Fig. \ref{xas_LaTiO3}.
The XAS spectra taken by different authors\cite{abbate_91,higuchi,haverkort} present some variations, possibly reflecting the presence of impurities in this compounds. The middle curve in Fig. \ref{xas_LaTiO3} show the latest work\cite{haverkort}. Both the crystal field corresponding to the cubic (dashed line in the figure) and to the orthorhombic structure were tested. In the latter case the spectrum becomes sensitive to the polarization direction. Two different polarizations were chosen for illustration purposes. The first one, labeled A (full lines in Fig. \ref{xas_LaTiO3}), coincides with a Ti-Ti direction for a Ti belonging to the next shell of Ti atoms. The second polarization direction, labeled B (dotted lines in the figure), is along the crystallographic $b$ direction of the compound. These two polarizations were chosen for their contrasting behaviors. The top curves in Fig. \ref{xas_LaTiO3} show the raw results of the multiplet code, the bottom after rescaling with S$_\text{coul}$=0.7, S$_\text{xtal}$=1.25 and S$_\text{soc}$=0.915. The raw result is calculated assuming a constant core-hole broadening of 0.25 eV. The scaled results are obtained with a constant core-hole broadening of 0.5 eV. The fits were done using the orthorhombic crystal field. Fitting starting from the cubic crystal field turned out more difficult in view of the positioning of the middle peak; the result of the rescaling using the parameters stated above but applied on the cubic crystal field are also shown in the figure (bottom dashed line).

\subsection{Monoclinic epidote}
The monoclinic epidote demonstrates the applicability of the code to low symmetry. Epidote has the generic formula Ca$_2$Al$_2$(Fe,Al)(SiO$_4$)(Si$_2$O$_7$)O(OH). The Fe$^{3+}$ in this compound is in a distorted octahedral environment formed by 6 oxygen atoms. XAS spectra were taken by Henderson {\it et al}\cite{henderson} on an epidote of unspecified composition for two polarizations, one in the approximate square-plane of the oxygen atoms and the other parallel to the axial compression, named $xy$ and $z$ polarization respectively. Many epidote structural data have been published for different Fe to Al ratio, but few differ substantially to one another when used as crystal field in the program. The latest neutron diffraction measurements\cite{gatta} was chosen for the results presented here (ICSD 168464)\footnote{Two Fe sites were identified in this study. The second one is shared with Al, but since the Fe occupancy of that site is only $1\%$, only one Fe site was considered.}.
 \begin{figure}
\includegraphics*[width=8cm,angle=0]{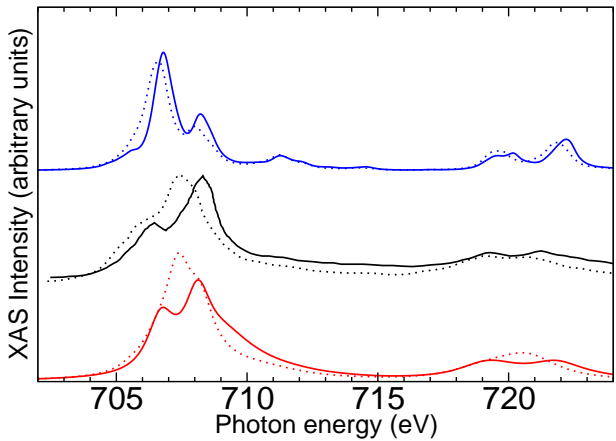}
\caption{(Color online) 
XAS spectra for Fe$^{3+}$ in monoclinic epidote in the ground state $2p^6 3d^5$. The dotted and full lines refer to respectively the xy and z polarization. Top (blue) lines: raw multiplet code results. Middle (black) lines: experimental results of Henderson {\it et al}\cite{henderson}. Bottom lines: code results with rescaled values as given in the text. \label{xas_epidote} }
\end{figure}
The XAS spectra for the transitions $2p^63d^5$ to  $2p^53d^6$ are plotted in Fig. \ref{xas_epidote}. The dotted lines refer to $xy$ polarization, the full line to the $z$ polarization. The middle curves in the figure are the measurements of Henderson {\it et al}\cite{henderson}, the top curves the raw output of the program and the bottom the results of rescaling. In this case the following scaling parameters were used: S$_\text{coul}$=0.7, S$_\text{xtal}$=1.8 and S$_\text{soc}$=0.915. For the bottom curves the broadening was increased from 0.55 eV to 1.2 eV from 708 eV onwards. \\

From the three cases presented, one can see that the general aspect and relative trend exhibited by the peaks for different polarizations are well reproduced by the raw output of the code. For the compounds studied here a good agreement with the experiment could be achieved with the screening of the Coulomb interaction, a slight adjustment of the spin-orbit coupling strength, and an enhancement of the crystal field strength. The rescaling of the crystal field is in effect an adjustment of the radial functions, i.e. their extent through the solid. The present scaling is therefore consistent with the modifications required to bring the point charge model closer to a more realistic crystal model.

\section{RIXS spectra \label{rixs}}
Very few high resolution RIXS spectra have been published to this date and even fewer relate to structures of very low symmetry. We have not been able to find a RIXS measurements where the effect of the low symmetry of the crystal field could be made unquestionably clear. We therefore demonstrate the RIXS capabilities of our code on two high symmetry examples for which experimental comparison is available, MnO and NiO. Furthermore, the case of distorted MnO is addressed, and results for a lanthanide compound are presented. \\

\subsection{Cubic MnO}
\begin{figure}
\includegraphics*[width=8cm,angle=0]{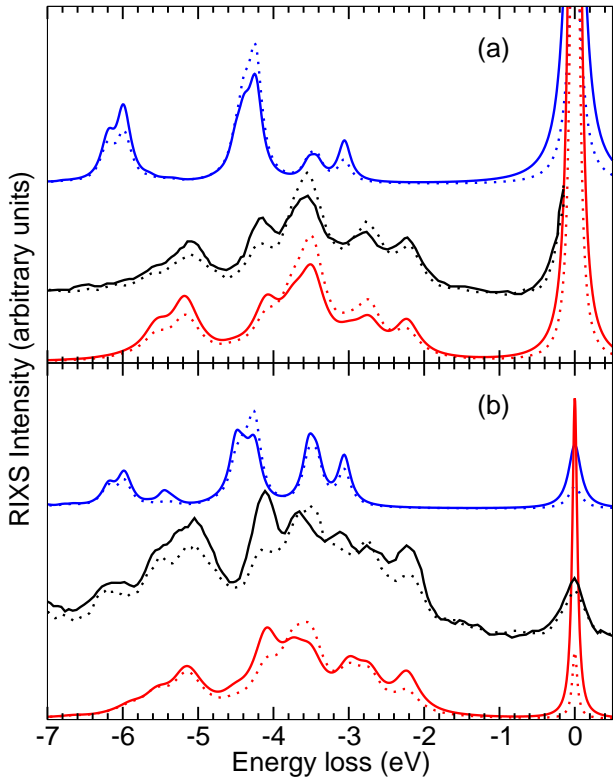}
\caption{(Color online) Cubic MnO: RIXS spectra for Mn$^{2+}$ with the $2p^63d^5$ ground state. Full lines correspond to the vertical polarization of the incoming photon, dotted lines to the horizontal one. Top (blue) curves: raw results of the code. Middle (black) curves: experimental results Ghiringhelli {\it et al.}\cite{ghiri}.  Lower (red) curves: results of the code with rescaling as given in the text. (a) and (b) are respectively taken at 0.7 eV and 3.7 eV above the L$_3$ edge.
}\label{MnOrixs}
\end{figure}

The Mn L-edge measurement in cubic MnO is a popular case for testing the resolution of the RIXS instrument. The valence state of Mn is Mn$^{2+}$ and corresponds to the ground state $2p^63d^5$. The RIXS process $2p^6 3d^5 \to 2p^5 3d^6 \to (2p^6 3d^5)^\star$ offers therefore insight into a non-trivial and rich multiplet structure. In a first place we assume that MnO has the cubic, NaCl structure, with a Mn-O nearest neighbor distance of 2.22 \AA. A series of measurements for several different incoming photon energies and two different polarizations were taken by Ghiringhelli {\it et al.}\cite{ghiri}. Fig. \ref{MnOrixs} (a) and (b) reproduces two sets of results, at 0.7 eV and 3.7 eV above the L$_3$ edge (label C and E in Ref. \onlinecite{ghiri}) respectively. 
In this figure, the curves measured or calculated with the incoming photon polarization horizontal, respectively vertical to the scattering plane are represented by dotted lines, respectively full lines. We refer the reader to Ref. \onlinecite{ghiri} for the description of the geometrical set up. The middle (black) lines in both panels are the experimental measurements\cite{ghiri}. The top (blue) lines are the raw results of the code, where a constant core-hole broadening of 0.2 eV and a constant final state broadening of 0.1 eV was taken. Although the raw spectra look somewhat shifted compared to the experiments, many of the features are already present. One notes in particular that the relative intensities for the two polarizations is reproduced. The bottom (red) curves in (a) and (b) are the spectra obtained after applying the scaling parameters S$_\text{coul}$=0.85, S$_\text{xtal}$=1.4 and S$_\text{soc}$=1.1. The rescaled curves are in excellent agreement with the experiments and compare favorably with the fitting done with the group-theory crystal field approach in Ref. \onlinecite{ghiri}. The latter fitting gave a value for the effective CXS of 1 eV. In order to get an estimate of the strength of the crystal field in the present approach, the ground state multiplet calculation is repeated with only 1 d electron, in the non-relativistic limit. In that case the $e_g$ and $t_{2g}$ levels clearly define CXS. Without scaling the raw result gives CXS=0.74 eV. When applying the scaling parameters CXS becomes 1.03 eV, an unsurprising result since the scaling parameters are deducted from the same fit as in Ref. \onlinecite{ghiri}.

\begin{figure}
\includegraphics*[width=8cm,angle=0]{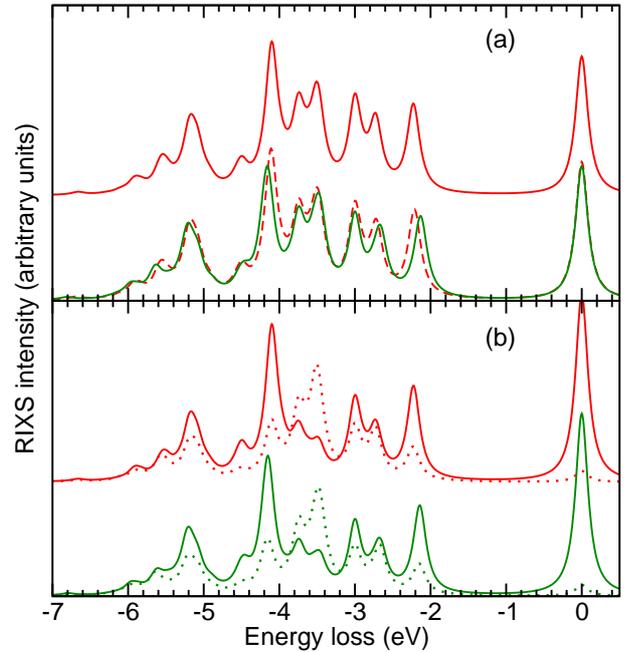}
\caption{(Color online) Calculated MnO RIXS spectra for Mn$^{2+}$ with the $2p^63d^5$ ground state; the same scaling parameters as in Fig. \ref{MnOrixs} were applied. The incoming photon has an energy 3.6 eV above the L$_3$ edge, and  $\bm \epsilon_{\text{in}}$ is in all cases along a Mn-O bond.
(a) Results with averages around $\bm b_\text{out}$; top (red) line: cubic MnO with $\bm b_\text{out}$ as in Ref. \onlinecite{ghiri_highres}; lower full (green) line: monoclinic MnO with $\bm b_\text{out}$ nearly perpendicular to $\bm \epsilon_{\text{in}}$; lower dashed (red) line: scaled cubic equivalent (see text).
(b) Results with given $\bm \epsilon_{\text{out}}$; top (red) lines for cubic, lower (green) lines for Mn1 in monoclinic MnO; full lines: $\bm \epsilon_{\text{out}} \parallel \bm \epsilon_{\text{in}}$; dotted lines: $\bm \epsilon_{\text{out}} \perp \bm \epsilon_{\text{in}}$ (only nearly perpendicular for monoclinic). 
}\label{MnOrixs0906}
\end{figure}

A high-resolution measurement in a slightly different geometry and taken at a different beamline was published three years later\cite{ghiri_highres}. Fig. \ref{MnOrixs0906}(a) (top red line) shows the calculated spectra for this particular experimental set up, with the same scaling parameters as used for Fig. \ref{MnOrixs}. 
The calculated spectra reproduces faithfully the features of the experimental measurement between -2 and -4 eV.

\begin{figure}
\includegraphics*[width=8cm,angle=0]{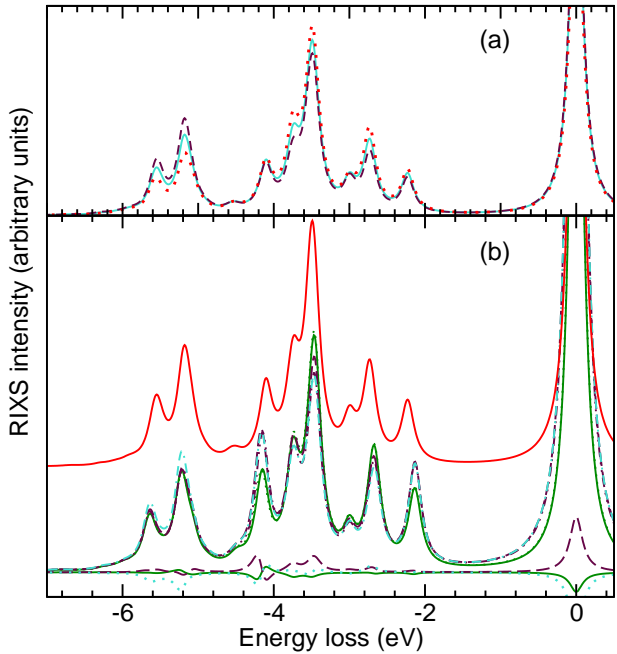}
\caption{(Color online) Calculated MnO RIXS spectra for Mn$^{2+}$ with the $2p^63d^5$ ground state; the same scaling parameters as in Fig. \ref{MnOrixs} were applied. The incoming photon has an energy 0.7 eV above the L$_3$ edge. (a) Cubic MnO; $\bm \epsilon_{\text{in}} \perp \bm \epsilon_{\text{out}}$, with neither $\bm \epsilon_{\text{in}}$ nor $\bm \epsilon_{\text{out}}$ along a bond.
(b) Top (red) curve: cubic MnO with $\bm \epsilon_{\text{in}}$ and $\bm b_\text{out}$ in two arbitrary directions, and with the respective two directions swapped. 
Middle curves: Monoclinic MnO with various $\bm \epsilon_{\text{in}}$ and $\bm b_\text{out}$ as given in the text (full green, maroon dashed, turquoise dash-dash-dotted lines), and with the directions of the first two swapped (green dotted, maroon dot-dot-dashed lines). Lower curves: differences between curves above, as described in the text.
}\label{MnOrixspol}
\end{figure}

As the present approach is very well suited to exploring the impact of different polarization directions on the spectra, we demonstrate in Fig. \ref{MnOrixs0906} (b) (top red curves) and Fig. \ref{MnOrixspol} (a) some of their effects in the cubic structure. To that purpose hypothetical spectra for selected incoming polarizations $\bm \epsilon_{\text{in}}$ and outgoing polarizations $\bm \epsilon_{\text{out}}$ are drawn.

The top (red) curves of Fig. \ref{MnOrixs0906} (b) show the calculated spectra of the cubic MnO for $\bm \epsilon_{\text{in}}$ along a Mn-O bond; $\bm \epsilon_{\text{out}}$  is either parallel (full line) or perpendicular (dotted line) to $\bm \epsilon_{\text{in}}$. As expected, the intensity of the elastic peak is weaker when $\bm \epsilon_{\text{in}} \perp \bm \epsilon_{\text{out}}$; also in this case the peaks at around -3.6 eV are enhanced, whereas those at -2.2 and -4.1 are severely reduced. 

Fig. \ref{MnOrixspol} (a) shows the effect of keeping $\bm \epsilon_{\text{in}} \perp \bm \epsilon_{\text{out}}$, but with neither directions along any symmetry axis of the structure. The calculations for the full (turquoise) and dashed (maroon) lines were performed with $\bm \epsilon_{\text{in}}$ chosen deliberately away from any bonds, namely in the (0.4,0.5,0.77) direction, with respect to the unit cell vectors. 
The outgoing polarizations $\bm \epsilon_{\text{out}}$ for those two lines are perpendicular to $\bm \epsilon_{\text{in}}$, being respectively in the (0.78,-0.62,0) and (0.35,0.690,-0.63) direction. For comparison, the dotted (red) curve shows $\bm \epsilon_{\text{in}} \perp \bm \epsilon_{\text{out}}$ with $\bm \epsilon_{\text{in}}$ along a bond. In this situation any outgoing polarizations perpendicular to it will produce the RIXS curve shown as the dotted line. When the incoming polarization is away from the symmetry axis, slightly different curves are produced even though the outgoing polarization is at $90^\circ$ from the incoming one.

\subsection{Low-temperature MnO}
MnO exhibits a phase transition\cite{goodwin_MnO} from paramagnetic to antiferromagnetic state at 118 K. A neutron scattering study\cite{goodwin_MnO} of MnO at 10 K shows that the distortions accompanying the phase transition transform the structure into a monoclinic one. Having established in the last section the sensitivity of RIXS to the different polarization directions in the cubic case, the question arises as to what extent these deformations could be picked up by RIXS. 

Three different Mn sites were identified in the monoclinic structure, all of them in distorted octahedral environments with 6 oxygen atoms as neighbors. The three distorted octahedra, although not exactly aligned, have similar orientations. In the cubic structure, drastic changes in the RIXS curves are observed for the outgoing polarization parallel and perpendicular to that of the incoming light, Fig. \ref{MnOrixs0906}(b), top (red) lines. Effects of a similar magnitude are expected to occur in the monoclinic structure. 

First the hypothetical spectra with selected incoming polarization $\bm \epsilon_{\text{in}}$ and outgoing polarization $\bm \epsilon_{\text{out}}$ are considered, Fig. \ref{MnOrixs0906} (b). The lower (green) curves of Fig. \ref{MnOrixs0906} (b) show the calculated spectra for Mn1 only, with $\bm \epsilon_{\text{in}}$ along the Mn1-O1 bond; the full line has been obtained for $\bm \epsilon_{\text{out}} \parallel \bm \epsilon_{\text{in}}$, the dotted line for $\bm \epsilon_{\text{out}}$ along the Mn1-O2 bond, which is nearly perpendicular to the Mn1-O1 direction.
The comparison between the top (cubic) and lower (monoclinic) curves of Fig. \ref{MnOrixs0906}(b) shows that the distortions are causing only minute changes to the spectra. One notes that the peak originally at -2.2 eV is shifted to the right by 80 meV, and that at 4.1 to the left by 50 meV, accompanied by some minor changes in height. 

The predicted spectrum for monoclinic MnO for a given $\bm \epsilon_{\text{in}}$ and a detection direction $\bm b_\text{out}$ is plotted in Fig. \ref{MnOrixs0906}(a) as the lower full (green) curve. For an optimal comparison with the cubic case, $\bm \epsilon_{\text{in}}$ has been taken along the Mn1-O1 bond, and $\bm b_\text{out}$ along the nearly perpendicular Mn1-O2 direction. The curve is the result of an average over the three Mn sites. In order to show the effect of distortions, the spectrum for the cubic structure has been recalculated with the point charges distances uniformly scaled down (dashed red line), so that the crystal field of the cubic structure matches the average of the distorted one. In effect, the cubic Mn-O distance $d$ has been changed so that $d^{-3}$ is the same as the average $\langle r^{-3}\rangle$ of the distorted octahedra, so that Mn-O distance is 2.212 instead of 2.22 \AA. The differences between the full (green) line and the (red) dashed line therefore exemplify the expected differences between the cubic and distorted structures. The same types of differences between the cubic and monoclinic can be noted as above in the discussion of Fig. \ref{MnOrixs0906}(b).

A more stringent test for the distortions can be worked out by noting that in the cubic structure, the same RIXS spectrum is obtained by exchanging the direction of the incoming polarization $\bm \epsilon_{\text{in}}$ with that of the unpolarized detection direction $\bm b_\text{out}$, for any chosen directions. An example is plotted in Fig. \ref{MnOrixspol} (b), top (red) curve. Two directions were picked randomly, $\mathbf{d}_1=(0.18,-0.40,-0.90)$ and $\mathbf{d}_2=(0.56,-0.31,-0.77)$ (with respect to the unit cell vectors). The same curve is produced by taking $\bm \epsilon_{\text{in}}$=$\mathbf{d}_1$ with $\bm b_\text{out}$=$\mathbf{d}_2$, and  $\bm \epsilon_{\text{in}}$=$\mathbf{d}_2$ with $\bm b_\text{out}$=$\mathbf{d}_1$. This is not the case for the distorted structure, for which slightly different curves are calculated when the directions are swapped. Some examples are shown as the middle curves in Fig. \ref{MnOrixspol} (b). The full and dotted (green) curves are produced with $\mathbf{d}_1$ along the Mn1-O1 bond, and $\mathbf{d}_2$ along the monoclinic crystallographic axis $a$; the full curve has  $\bm \epsilon_{\text{in}}$=$\mathbf{d}_1$ with $\bm b_\text{out}$=$\mathbf{d}_2$, the dotted curve $\bm \epsilon_{\text{in}}$=$\mathbf{d}_2$ with $\bm b_\text{out}$=$\mathbf{d}_1$. The difference between these two curve is plotted as full (green) line in the lower part of the figure. A greater difference is observed by taking $\mathbf{d}_1$ along $a$, and $\mathbf{d}_2$ perpendicular to the crystallographic directions $a$ and $b$, which we call $c'$. The relevant curves are drawn as (maroon) dashed and dot-dot-dashed, with the difference plotted as dashed line. The middle (turquoise) dash-dash-dot curve was calculated with $\bm \epsilon_{\text{in}}$ along $b$, and $\bm b_\text{out}$ along $a$. The difference between this spectrum and one with $\bm \epsilon_{\text{in}}$ along $c'$ and $\bm b_\text{out}$ along $a$ is shown in the lower part of the figure as the (turquoise) dotted line. Here the difference is greater in the lower part of the spectra. It is expected that those tiny differences, in this case smaller as $10\%$ of the intensity of the curve,  will be very difficult to detect experimentally.

\subsection{Cubic NiO}
\begin{figure}
\includegraphics*[width=8cm,angle=0]{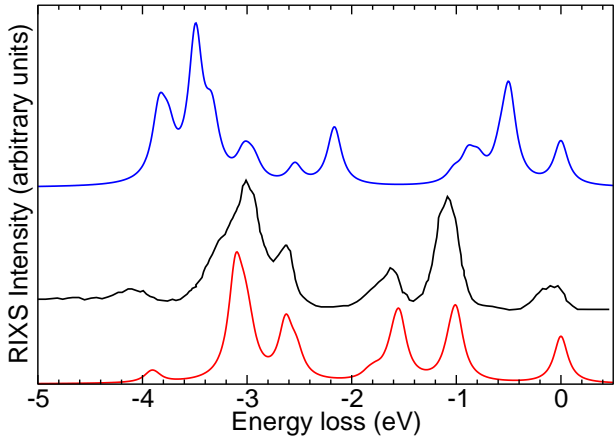}
\caption{(Color online) RIXS spectra for Ni$^{2+}$ with the $2p^63d^8$ ground state in NiO. Top curve (blue): raw results of the code. Middle curve (black): Experimental result of Ghiringhelli {\it et al.}\cite{ghiri_highres}. Bottom curve (red): result of the code after scaling (see text).
}\label{NiOrixs}
\end{figure}

Another example where a simple crystal field model can be applied is at the Ni L-edge of NiO\cite{ghiri_NiO}. The high-resolution experiment of Ghiringhelli {\it et al.}\cite{ghiri_highres} is reproduced as the middle (black) curve in Fig. \ref{NiOrixs}.
Adopting the set up described in Ref. \onlinecite{ghiri_highres} the calculations were carried assuming cubic symmetry for NiO with the shortest Ni-O distance at 2.09 \AA. Ni in NiO as the valance Ni$^{2+}$ and the ground state is therefore $2p^6 3d^8$. The top (blue) line in Fig. \ref{NiOrixs} is the raw result of the code. The bottom line (red) is obtained by applying the scaling parameters S$_\text{coul}$=0.7, S$_\text{xtal}$=2.0 and S$_\text{soc}$=0.915. A constant core-hole broadening of 0.1 eV and final state broadening was applied to both calculated curves. Although again the raw curve contains the right features with the correct hierarchy of peaks, a fairly strong rescaling of the crystal field strength had to be applied in that case to bring the curve in good agreement with the experiment. The CXS value can be obtained in a similar fashion as for MnO by placing a single $d$ electron in the $3d$ shell and performing the calculation in the non-relativistic limit. The raw multiplet result gives an unscaled CXS of 0.51 eV and a scaled CXS of $1.03$ eV. The latter value is in complete agreement with the group-theory approach of Ref. \onlinecite{ghiri_NiO}.

\subsection{Dy phthalocyanine}
Lastly, we use our code to predict the crystal-field effects on the RIXS spectrum of a lanthanide compound, the ``double-decker'' phthalocyanine complex\cite{katoh_Dy}, with formula C$_{64}$H$_{32}$Dy$_1$N$_{16}$. The multiplet levels for this structure have been determined by Ishikawa {\it et al} by combining NMR measurements and ligand-field parameters modeling\cite{ishikawa}, thereby providing a point of comparison with the ground state multiplets calculated by our code. In the selected structure a trivalent Dy atom sits between two planar ligands composed of the N, C and H atoms. Dy$^{3+}$ is in the ground state $3d^{10}4f^9$. 

\begin{figure}
\includegraphics*[width=8cm,angle=0]{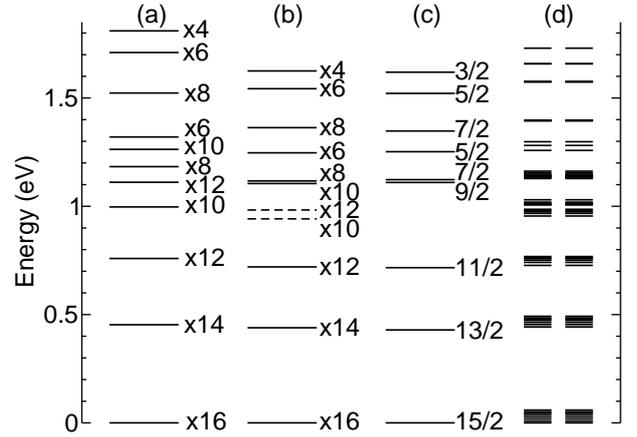}
\caption{Low-lying multiplets for Dy$^{3+}$ in the $3d^{10}4f^9$ ground state. (a) and (b) Calculated levels in zero crystal field with respectively S$_\text{coul}$=S$_\text{soc}$=1, and   S$_\text{coul}$=0.85, S$_\text{soc}$=0.95, and associated multiplicity. (c) Ion spectroscopy experiments\cite{nist} and associated $J$ value. (d) Calculated multiplets in crystal field (see text) with S$_\text{coul}$=0.85 and S$_\text{soc}$=0.95.
}\label{Dy_levels}
\end{figure}

Starting without a crystal field, the energy levels for Dy$^{3+}$ can be calculated and compared with those reported in the literature for Dy IV\cite{nist}. The latter are plotted in column (c) of Fig. \ref{Dy_levels}, with their corresponding $J$ value.
The columns (a) and (b) show the lowest levels and their multiplicity, respectively with no scaling (a) and by reducing the intra-coulomb interaction by S$_\text{coul}$=0.85 and the spin-orbit coupling by  S$_\text{soc}$=0.95 (b). The raw result of the code is in excellent agreement with the measurements (c) for the first levels, with a small shift upwards for the higher plotted values. The match is further improved by the application of rescaling. Two of the calculated levels (dashed lines in the figure) were not reported in Ref. \onlinecite{nist}; the other levels are in one to one correspondence with the measurements.

As starting point for the crystal field charges, the oxidation states N$^{-3}$, C$^{+4}$ and H$^{+1}$ are a poor choice in view of the non-ionic character of this compound. A DFT calculation indicated that the charges are close to being a tenth of the oxydation state values\footnote{Using Hirshfeld's spatial partitioning of the charge density}. As an alternative to scaling the charges down, one can set the crystal field scaling parameter S$_\text{xtal}$ to 0.1. Column  (d) show the first 54 levels calculated with this crystal field and the scaling parameters S$_\text{coul}$=0.85 and S$_\text{soc}$=0.95. One notes that the $16$-fold degenerate ground state has split under the influence of the crystal field in a series of low-lying, two-fold degenerate states, all very close to zero. Ishikawa {\it et al} find the first level at 0.004 eV and the last at 0.068 eV; the present code is in close agreement with respectively 0.008 and 0.059 eV.

\begin{figure}
\includegraphics*[width=8cm,angle=0]{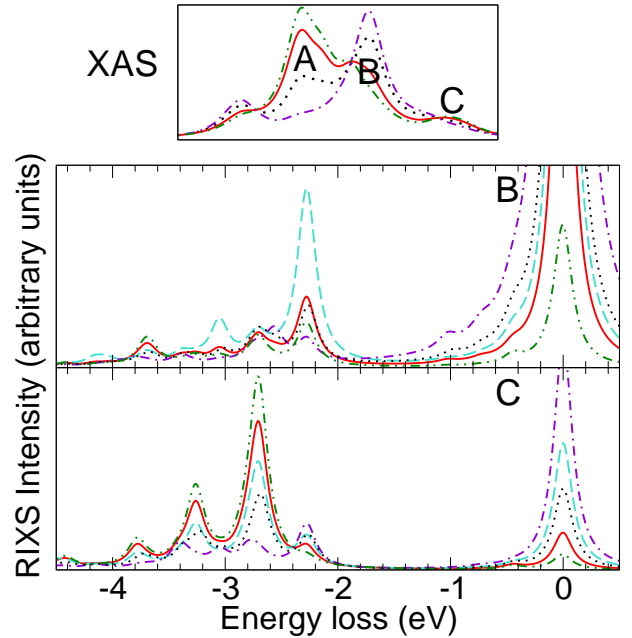}
\caption{(Color online) Calculated XAS and RIXS spectra for Dy$^{3+}$ with the $3d^{10}4f^9$ ground state in Dy phthalocyanine. XAS: with polarization along $a$ (full red line), $c$ (dash-dotted violet line), and $C_4$ axis (dash-dot-dotted green line); (black) dotted line are calculated in the absence of crystal field. RIXS: full (red) lines: $\bm \epsilon_{\text{in}} = \bm b_\text{out} \parallel a$, dashes (turquoise): $\bm \epsilon_{\text{in}} \parallel a$, $\bm b_\text{out} \parallel c$, dash-dots (violet): $\bm \epsilon_{\text{in}} \parallel c$, $\bm b_\text{out} \parallel a$, dash-dot-dots (green): $\bm \epsilon_{\text{in}} = \bm b_\text{out}$ along $C_4$ axis of the molecule; the dotted (black) lines have been produced with  $\bm \epsilon_{\text{in}} = \bm b_\text{out}$ in the absence of crystal field, and have been scaled down by a factor $8$.
}\label{Dy_xas_rixs}
\end{figure}

 The XAS and RIXS spectra for the Dy phthalocyanine complex calculated with these scaling parameters are plotted in Fig. \ref{Dy_xas_rixs}. 
The XAS in Fig. \ref{Dy_xas_rixs} shows the M$_5$ part of the spectra and indicates the energies at which the RIXS spectra were calculated. The following incoming polarization directions were used: along the crystallographic axis $a$ (full red line), along $c$ (dash-dotted violet line), and along the $C_4$ axis (green dot-dot-dashed line) of the molecule. The (black) dotted line shows the result in absence of crystal field. The energies B and C are respectively at $2.3$ and $4.9$ eV from the point A. The RIXS spectra were calculated for various incoming polarizations $\bm \epsilon_{\text{in}}$ and detection directions $\bm b_\text{out}$. The full (red) lines were generated for $\bm \epsilon_{\text{in}} = \bm b_\text{out}$ both along the crystallographic axis $a$. The dashed (turquoise) curves also had $\bm \epsilon_{\text{in}}$ along $a$, but $\bm b_\text{out}$ along $c$, while the dash-dotted (violet) lines took $\bm \epsilon_{\text{in}}$ along $c$, and $\bm b_\text{out}$ along $a$. Setting $\bm \epsilon_{\text{in}} = \bm b_\text{out}$ along the $C_4$ axis of the molecule gives the dash-dot-dotted (green) curves. The RIXS intensity with no crystal field (black dots) were scaled down by a factor $8$ for an optimal comparison with the other curves. This factor reflects the respective degeneracy of the ground state levels with and without the application of the crystal field. The lowest lying states are absorbed by the elastic peaks, but the states between $0.4$ and $1$ eV are visible. Another feature is the predominance of the peak at $2.3$ eV at energy B, then of the peak at $2.7$ eV at energy C, for the incoming polarization along $a$. Although it is likely that the chosen scaling parameters would need some adjusting, testing with parameters indicated that the major effect on the RIXS spectra of Fig. \ref{Dy_xas_rixs} will appear to produce a shift of the peaks between $2$ and $4$ eV along the energy axis, rather than drastic changes in polarizations behavior.  \\

In this section RIXS calculations on cubic MnO and NiO were tested against measurements. The raw curves exhibited the correct relative behavior, and very good agreement with experiments could be achieved by applying scaling parameters. The level of changes in the spectra caused by distortions could be evaluated for the low temperature MnO. Finally, the code was applied on an $f$ electron compound, Dy phthalocyanine. After establishing by comparisons with measurements that reasonable multiplets levels are obtained for the ground state, RIXS predictions were made.

\section{Conclusion and Outlook \label{conclusions}}
In summary, we have described an easily accessible approach for the
computation of multiplet spectra arising from soft X-ray 
spectroscopies of narrow band solids. 
The required input data is 
\begin{itemize}
\item  The element being probed (for example Ti) 
\item  The core and valence configuration of the ground state for the shells involved in the transitions (for example $2p^6$ $3d^2$)
\item  The crystal field as a list of positions and charges
\end{itemize}
The charges can be viewed as semi-empirical, with formal charges or oxidation states as a starting guess.
The central field part of the present theory is done with a Dirac relativistic 
implementation for a simple density functional, and yields the spin-orbit 
splitting from first principles.
The electron-electron interaction is calculated explicitly for the electrons of the open shells. The crystal field is obtained as the resulting electrostatic field generated by the surrounding point charges. Any crystal field of arbitrary and unspecified symmetry can therefore be entered without expert knowledge of group 
theory notation. Moreover, this approach allows the polarization directions to be given in a straightforward manner with respect to the positions of the crystal field ions. Within the limitations of the model and the scope of the approximations made, the method constitutes a ``first principles'' approach for calculating multiplets states as well as XAS and RIXS spectra. In the latter cases the core-hole lifetime is also required as an input.

Our examples show that the raw, ``first-principle'' results provide a good estimation of the multiplet structure; the correct experimental trends are reproduced in those experiments or part of the spectrum where the crystal field plays a dominant role. The model of the single atom in an ionic environment limits however the agreement with experiments. This can be corrected to some extent by the introduction of semi-empirical scaling parameters that account for screening effects, hybridization and coupling to energy bands that are not originally within the scope of the model. The scaling parameter for the intra-atomic Coulomb interaction varies typically between 0.7 and 1; a 10$\%$ adjustment of the spin-orbit coupling has sometimes been necessary; in the examples discussed in this paper the value of crystal field scaling parameter has ranged from 1.0 and 2.0. These corrections to the raw result reflect the fact that in a bonded solid, the radial functions extend further than anticipated by the neutral atom calculation. The crystal field correction may also include a global adjustment to the point charge values.

Although the use of scaling parameters leads to excellent fits to experiments, it is necessary to build up the model and include non-local contributions into the spectra. The crystal field model for any symmetry constitutes the first step towards a more complete approach to multiplets for core spectroscopies.

\begin{acknowledgments}
We thank our present and former colleagues of the ADRESS beamline at 
the Swiss 
Light Source: Thorsten Schmitt, Justina Schlappa, Kejin Zhou, Claude Monney and 
Luc Patthey, for valuable inputs during our discussions. 

We gratefully acknowledge D. D. Koelling for the Dirac central field 
radial equation solver. 

We also thank C. Dallera, C. McGuiness and 
J. Nordgren for stimulating discussions and 
encouragement.

This work has been funded by SNF grant 200021-129970.
\end{acknowledgments}

\bibliography{mybib} 
\end{document}